\documentclass{amsart}
\usepackage{epsfig}
\usepackage[cp1250]{inputenc}
\usepackage{graphicx}

\theoremstyle{definition}

\theoremstyle{remark}

\numberwithin{equation}{section}

\begin{document}

\title[Gamow's bicycle]{Gamow's bicycle:\\The Appearance of Bodies at Relativistic Speeds and Apparent Superluminal Velocities}%
\author{Andrzej Nowojewski}%
\address{Mansfield College,University of Oxford, Oxford OX1 3TF, UK}%
\email{vanshar@hot.pl}%


\begin{abstract}
A human creates an image basing on the information delivered by photons that arrived at his retina simultaneously.
Due to finite and constant velocity of light these photons left the moving body at different times, since not all
points of the body are equidistant. In other words its image represents the body as it was in several different
times i.e. it is distorted and does not correspond to its real appearance. The useful experimental arrangement is
set and then used to derive the general expression that transforms two-dimensional stationary shapes to their
apparent forms, which could be photographed once they are set in motion. It is then used to simulate the so-called
Gamow's bicycle combined out of circles and straight lines. The simulation outlines two important aspects of
bike's motion: apparent distance of two points and apparent velocity which are then discussed thoroughly. It is
found that the approaching body is elongated and its apparent speed is greater than its real one (under certain
conditions can exceed the speed of light), whereas the receding one is contracted (but not in a matter of Lorentz
contraction) with the speed smaller than the real one. Both the apparent length and speed tends to a certain limit
when time tends to $\pm\infty$. The change of both parameters takes place in the vicinity of the nearest approach
to the observer and is more rapid when the velocity greater and the distance is smaller. When the moving vertical
rod is seen at right angle, its total apparent length is Lorentz contracted, however its interior is still
distorted. At the same conditions apparent velocity of a point equals its real one. Finally it is proven that not
only the apparent geometrical shape changes as the body moves but also its color according to Doppler shift.
\end{abstract}
\maketitle
\section{Introduction}

\par Once the revolutionary Einstein's Theory of Relativity was understood, science educators began publishing books
and articles that gave an average person the chance to comprehend
these new ideas. One of such books, George Gamow's "Mr. Thompkins in
the Wonderland"$^{[1]}$, became a bestseller. The main character
visits a city where the speed of light just slightly exceeds the
velocity of a bicycle. As he sees an approaching cyclist, two most
controversial (to a non-physicist) effects of Special Relativity -
length contraction and time dilatation - are displayed to the
shocked traveller. But is it really the picture Mr. Thompkins would
see? Gamow in his considerations has not taken into account the fact
that Mr. Thompkins perceives the bicycle thanks to individual
photons, which travel with finite velocity. In this paper we develop
a method of simulating the appearance of moving, two-dimensional
objects, present the results in the example of Gamow's bicycle and
provide their detailed discussion. Some outcomes of this paper have
already been published$^{[2]}$.

\par Mr. Thompkins's brain creates an image of a moving bicycle
basing on information that was delivered by photons, which arrived on his retina at the same time. This, however
does not mean that these photons were emitted by the bicycle at one instant too. We have to remember that light
travels at a finite velocity of $c$, which according to special theory of relativity is constant for all
observers. Accordingly, photons emitted from rear light of approaching bicycle will reach Mr. Thompkins later than
photons from front light. If photons from both lights reach the observer simultaneously, the light from the rear
lamp must have been emitted earlier. That is why the image will appear distorted and elongated, since the bike was
further away at earlier times (we will call the phenomenon by the name of \textit{retardation effects}). An
analogous situation occurs in astronomy. We can observe celestial bodies as they were $3 \cdot 10^6$ years ago
(e.g. M32 Galaxy of Andromeda) and 5000 years ago (e.g. M1 Crab Nebula) at the same time. The closer the object
is, the more recent image we receive: M32 is $3 \cdot 10^6$ light years whilst M1 is only 5000 light years away.
It is a result of finite velocity of propagation of light and information. We do not experience such phenomena in
our every day life because it only becomes apparent in extreme situations: either when we deal with great
distances (e.g. astronomy) or great velocities (e.g. Gamow's bicycle).

\par Gamow's book presents either unintentionally or deliberately (in order to simplify complex
matters of Einstein's theory of relativity) incorrect statements concerning the visual appearance of bodies in
motion. In fact, it only represents the popular conviction of physicists that Lorentz contraction could be
visually observed. Even Lorentz and Einstein themselves suggested this statement to be true$^{[3]}$. The first
attempt to contradict the general belief was made by Lampa in 1924$^{[4]}$. Unfortunately it was overlooked by the
international community and it was not until Terrell in 1959$^{[3]}$ when heated discussion arouse on the topic,
both in professional journals as well as popular-science magazines$^{[5][6]}$. Terrell showed that Lorentz
contraction cannot be observed visually and rapidly moving, distant or small (i.e. subtending small solid angles)
bodies will instead appear rotated. It was not long before many more independent investigations of the phenomenon
were launched. Analysis of a moving sphere showed that it appears to be rotated and its circumference will always
(regardless of the distance or its radius) be circular$^{[7][8]}$. Analogous results were obtained for the case of
a moving cube$^{[9]}$. It was later on discovered that the interpretation of the results and some conclusions in
early publications were inaccurate. The most comprehensive revision of the effect was done by Mathews and
Lakshmanan$^{[10]}$. It was shown that understanding the retardation effects as an apparent rotation could lead to
a common belief that moving bodies are in fact rotated, which results in a number of absurdities in the so-called
"train paradox"$^{[11]}$. Apparent rotation is grossly misleading: change in apparent volumes and distortion of
cross sections of moving bodies could not be explained by a simple rotation but by a combination of shears,
extensions or contractions parallel to the direction of motion (under certain conditions a sphere may even appear
to be concave$^{[12]}$). Finally Scott and Viner$^{[13]}$ pointed out correctly that under particular conditions
Lorentz contraction could be observed directly. Unfortunately by the time these errors where corrected, several
textbooks on special relativity$^{[14][15]}$ as well as instructional articles$^{[16]}$ were published with these
mistakes. Fortunately more recent papers$^{[17]}$ seem to recognize the source of the problems. In the meantime,
considerations of the apparent, geometrical shape of moving bodies were supplemented by calculations of light
intensity and spectra changes$^{[18][19][20]}$ that resulted in more complete description of the retardation
effects.

\par In this paper we will derive an expression
that transforms two-dimensional, stationary objects into shapes that could be observed while these bodies are at
motion. We present how the operation works in the example of Gamow's bicycle, which is then thoroughly analyzed
under two main aspects (its apparent speed and the apparent distance of two arbitrarily chosen points). We will
use conventional notation: Lorentz factor: $\gamma=1/\sqrt{1-(v/c)^{2}}$ and $\beta=v/c$. For the sake of
simplicity, we assume in all simulations that $c=1$. All figures and simulations were created with the programme
\textit{Mathematica 4.1}.

\section{Finite velocity of photons}

\par Let us consider two cartesian frames of reference $S'$ and $S$,
moving relative to each other at the speed of $v$, parallel to $x$ and $x'$ axes in the positive direction of
$x$-axis . The origins of both systems and all axes $(x,y,z)$ coincide at $t=t'=0$. For the purpose of this paper
we shall only consider two-dimensional figures stationary in $S'$ and satisfying the equations $F(x,y)=0$, $z=d$
in $S'$. According to special theory of relativity $F$ will be contracted in $S$:
\begin{equation}\label{2.1} F(\gamma(x-\beta ct){},y)=0 \end{equation}
The picture below illustrates how the Gamow's bicycle really looks like in $S$ and $S'$.
 \begin{figure*}[h]
 \centering
 \epsfig{file=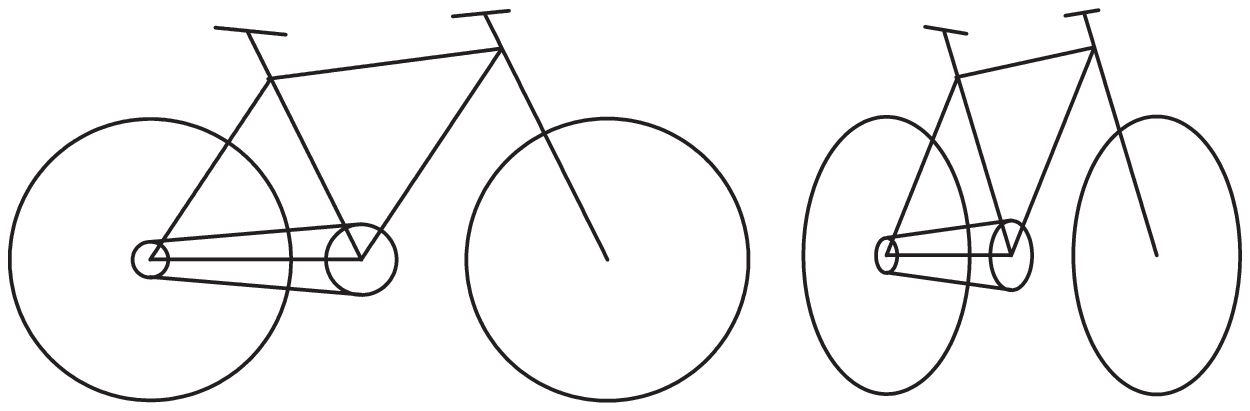, width=11cm}
 \caption{A picture showing Gamow's bicycle at rest (left) and a contracted bicycle, moving with the velocity $\beta=0.8$ (right).}
 \end{figure*}

 \begin{figure*}[p]
 \centering
 \epsfig{file=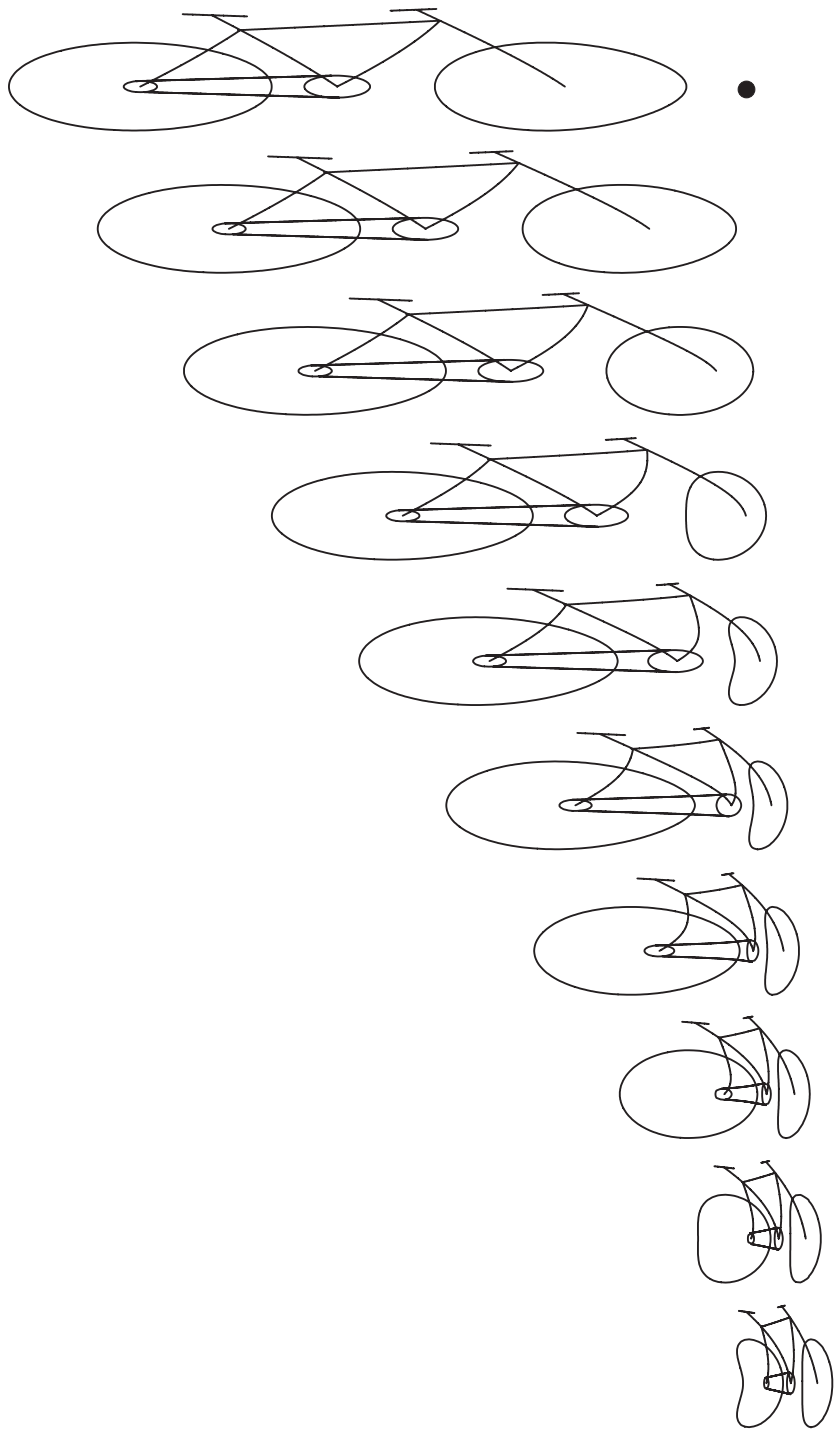, width=11cm}
 \caption{A sequence of 10 photographs of a Gamow's bicycle moving with velocity $\beta=0.8$ to the right. Pictures are
  separated with the same time intervals: first was taken at $t_{0}=-4$, the last $t_{0}=14$ (the actual center of the front
  wheel is at the least distance from an observer at time $t_{0}=0$). The observer (indicated here by a black dot)
  is located at a distance equal to half the diameter of wheels from the plane of motion.}
 \end{figure*}

\par Before we proceed in deriving the transformation let us think about choosing an appropriate type of observation. The choice
of an observer was generally not discussed by the researchers with few exceptions$^{[9]}$. At first thought it may
not be obvious that Mr. Thompkins or, in general, a human with a naked eye are not ideal observers for such
considerations. An image created in a brain is a projection on a plane that is inclined at some angle to the
$x$-axis in system $S$, because one usually turns one's head towards the moving object during the process of
observation. The choice of the angle of inclination is not as easy as it may appear in case of rapidly moving
objects, since they become distorted and there is no easy way to tell where the geometrical center is located.
Apart from that, humans posses a depth vision thanks to a pair of eyes. Three dimensional vision is created by
combining two images from both eyes that do not match each other exactly (because of the distance separating human
eyes). Again, in our case it causes more harm than good. Because of the separation, at any instant of time, one
eye will always be closer to the object. Because of that it will receive an earlier image of the object. In that
case images from both eyes will not only be shifted but indeed will not match each other at all (for that matter
it is questionable whether the approach used by Kraus $^{[18]}$ to describe the appearance of a rapidly moving
sphere seen by a human observer was useful)! Even if we developed a method to simulate such images it would be
useless since no qualitative observations could be done. Therefore we have to devise a system that allows us to
make some measurements. Let us imagine that the $x$-axis is a non-transparent barrier with a small aperture with a
shutter located exactly at the origin of $S$ (the point with coordinates $(0,0,0)$). We also assume that the
entire plane at distance $-l$ ($l>0$), parallel to the $XY$ plane is a very sensitive film (i.e. with infinite ISO
number). This arrangement will alow us to visualize the actual appearance of a moving object without committing
our attention to an implementation of complicated projections that would further distort the image. Now, we assume
that a group of photons from the self-luminous body $F$ (see Eq. (2.1)) arrives at an aperture at time $t_{0}$.
The shutter is opened at that moment to allow the photons to arrive later on the film. The distance $l$ is in fact
irrelevant because it only enlarges the picture isotropically. We may then choose convenient, special case $l=d$
and not discuss it further. This way, we finally obtain a photograph of $F$ at the instant of observation $t_{0}$.
Owing to equation (2.1) we know the location of every point of $F$ as a function of time. An arbitrary chosen
photon emitted from point with coordinates $(x,y,d)$ at time $t$ will travel the distance
$\sqrt{x^{2}+y^{2}+d^{2}}$ before it arrives at the origin at time of observation $t_{0}$. We can write down a
simple equation of motion for that photon:
\begin{equation}\label{2.2}
  c(t_{0}-t)=\sqrt{x^{2}+y^{2}+d^{2}}
\end{equation}
We now have two equations: (2.1) which relates the location of emission points in space to the time $t$ and (2.2)
which links the location and emission time $t$ with the time of observation $t_{0}$. The observer cannot be sure
whether the photons were emitted at one instant or not. He only knows that photons arrived simultaneously at
$t_{0}$. We will therefore calculate $t$ from (2.2) and substitute this expression to (2.1):
\begin{equation}\label{2.3}
    F(\gamma(x-\beta(ct_{0}-\sqrt{x^{2}+y^{2}+d^{2}})){},y)=0
\end{equation}
This is the expression that could be used to find out how the figure $F$ would appear at time $t_{0}$ when moving
with velocity $v$. It is clear that only $x$ coordinate is affected, i.e. retardation effects work only in the
direction of motion. That is why interpreting it as an apparent rotation of an object in motion (the so-called
Terrell-Penrose rotation) is unreasonable - any rotation changes coordinates in at least two dimensions. In order
to find out how this expression works in practice let us consider Gamow's bicycle from Fig. 1. It comprises out of
simple, geometrical figures (4 circles and 10 lines) so applying (2.3) should not be a problem.

\par What does the Fig. 2 actually transpire? The first striking observation is that the apparent shape of a bicycle is
nowhere near the shape of a contracted one (Fig. 1). An approaching vehicle is elongated, whereas a receding one
is contracted (but not in a matter of Lorentz contraction). The change takes place when a bicycle goes past the
observer. At that time every part of the bicycle is deformed and its view is generally distorted. At one point
each wheel may even resemble a banana or a croissant. If we look closer at the movement of both axles we notice
that their speed is different and it is changing, despite that the velocity of the bicycle is constant. The
velocity of an approaching axle is greater than that of the receding one. In following sections we will explain
these phenomena.
\section{Apparent length}

In Fig. 2 we can clearly see that the apparent distance between two points (e.g. front and rear axles) changes in
time. This section is an extension of Weinstein's paper$^{[21]}$ that first dealt extensively with this topic. We
can investigate this phenomenon by considering two points (e.g. both ends of a rod). In $S'$ their coordinates
are: $x_{1}'=0$, $x_{2}'=L$ and are separated by the distance $x_{2}'-x_{1}'=L$. In $S$, however, we arrive with
following equations of motion thanks to Lorentz transformation:
\begin{equation}
    x_{1}'=\gamma(x_{1}-\beta ct)\Leftrightarrow x_{1}=vt,
\end{equation}
\begin{equation}
x_{2}'=\gamma(x_{2}-\beta ct)\Leftrightarrow x_{2}=L/ \gamma + vt,
\end{equation}
hence, the distance separating these two moving points is contracted:
\begin{equation}
    L'=x_{2}-x_{1}=L/ \gamma.
\end{equation}
Now we can calculate the distance between these points that will actually be seen. Let us assume that they are
moving along the line parallel to the $x$-axis at a distance $d$ from it. Light impulses emitted at times $t_{1}$
from rear and $t_{2}$ from front point reach observer at $t_{0}$. Their equations of motion are:
\begin{equation}
    c(t_{0}-t_{1})=\sqrt{x_{1}^{2}+d^{2}},
\end{equation}
\begin{equation}
    c(t_{0}-t_{2})=\sqrt{x_{2}^{2}+d^{2}}.
\end{equation}
Now we calculate $t_{1}$ and $t_{2}$ from (3.1) and (3.2):
\begin{equation}
    t_{1}=x_{1}/v,
\end{equation}
\begin{equation}
    t_{2}=\frac{x_{2}-L/ \gamma}{v}
\end{equation}
and implement them in (3.4) and (3.5):
\begin{equation}
    c(t_{0}-x_{1}/v)=\sqrt{x_{1}^{2}+d^{2}},
\end{equation}
\begin{equation}
    c(t_{0}-\frac{x_{2}-L/ \gamma}{v})=\sqrt{x_{2}^{2}+d^{2}}.
\end{equation}
On solving these two equations we obtain:
\begin{equation}
    x_{1}=t_{0}v\gamma^{2}-\beta \gamma \sqrt{d^{2}+(t_{0}v\gamma)^{2}},
\end{equation}
\begin{equation}
    x_{2}=L\gamma+t_{0}v\gamma^{2}-\beta \gamma \sqrt{d^{2}+(L+t_{0}v\gamma)^{2}}.
\end{equation}
Finally the observed length equals:
\begin{equation}
    L_{0}=x_{2}-x_{1}=L\gamma-\beta \gamma(\sqrt{d^{2}+(L+t_{0}v\gamma)^{2}}-\sqrt{d^{2}+(t_{0}v\gamma)^{2}}).
\end{equation}
 \begin{figure*}[h]
 \centering
 \epsfig{file=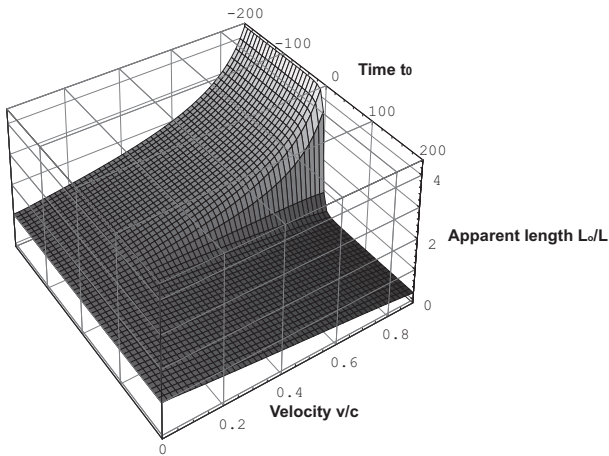, width=8cm}
 \caption{The graph presents the dynamics of apparent length as a function of velocity and observer's time.}
 \end{figure*}
It obviously differs very much from an ordinary Lorentz contracted length (3.3), it depends not only on velocity
$v$ and stationary separation $L$ but also on the distance $d$ and the time of observation. We will now analyze
dependance of $L_{0}$ on velocity and distance as a function of time.
\par From an analysis of Fig. 2 we already know that an approaching object is elongated whereas a receding one is
contracted. We also suspect that $L_{0}\approx L$ when $v\ll c$ since we do not encounter any such phenomena in
every day life. In fact if we look closely at (3.12) we will see that the expression tends to $L$ when
$v\rightarrow 0$. To see how $L_{0}$ behaves when $v$ changes we turn to Fig. 3. When $v$ is small $L_{0}$ tends
indeed to $L$, on the other hand when $v \approx c$, $L_{0}$ could be several times greater or smaller than $L$.
We can also see that the graph is consistent with Fig. 2. An approaching rod seems elongated several times and a
receding one contracted by roughly the same magnitude. The change takes place near the observer. This contraction
is more rapid when velocity is closer to $c$. We also observe another interesting fact. The apparent length
approaches a certain limit when time tends to infinity. We calculate it by putting $t_{0}\rightarrow \pm \infty$
in (3.12):
\begin{equation}
    L_{0\pm\infty}=L\gamma(1\mp\beta).
\end{equation}
This equation resembles a relativistic Doppler shift (it was first noticed by Mathews and Lakshmann$^{[10]}$)!
This fact should not come to us as a surprise considering, that the contraction or elongation of light waves is
caused by exactly the same fact (i.e. finite and constant velocity of light). It is, however necessary to draw the
following distinction: Doppler shift is real and retardation effects discussed in this paper are only apparent. In
addition it is worthy to point out that under special circumstances Lorentz contraction could be observed visually
(contradictory to some early statements$^{[3]}$ ). It is so, when both ends of the apparent rod are at the same
distance$^{[13]}$  - photons from each end travel the same time to the observer (there is no retardation). In our
case it happens at time $t_{0}=\sqrt{L^{2}/4\gamma^{2}+d^{2}}/c-L/2v\gamma$.
 \begin{figure}[h]
 \centering
 \epsfig{file=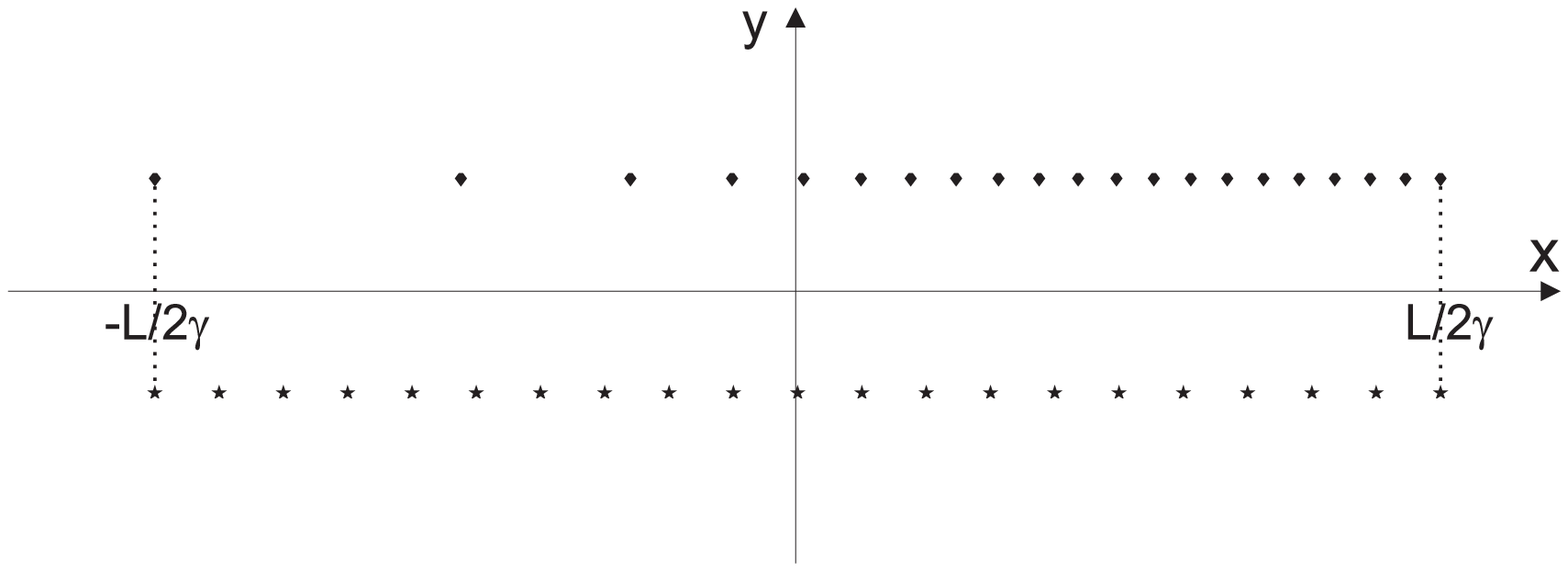, width=8cm}
 \caption{Two sets of 21 points, each representing equally distributed features on the moving vertical rod. Lower set
 shows the real appearance of the moving rod at time $t=-L/2v\gamma$ (emission time) and the upper one, the observable appearance
 at time $t_{0}=\sqrt{L^{2}/4\gamma^{2}+d^{2}}/c-L/2v\gamma$ (observer's time), i.e. when both ends of the rod were equidistant from the
 origin (both rods move along a line, parallel to x-axis at the distance $d=L/5\gamma$ with $v=0.9c$, here they have been
 separated for the sake of clarity). We can clearly see that the total lengths of both rods are Lorentz-contracted, however
 the interior of the observable rod (upper) is distorted.}
 \end{figure}
 However this statement is not true in general. We have to remember that merely rod's ends, and not the interior, are equidistant from
 the origin. In other words, photon emitted from the point that is located within the rod will travel shorter
 distance before reaching the observer. Therefore retardation effects will still apply and the interior of the rod
 would be distorted. We can fully observe this effect in Fig. 4. Unfortunately Scott and Viner failed to recognize
 this fact in their criticism of Terrell. The interior distortion is only negligible when the rod's length is far
 smaller than the distance from the origin, i.e. $d\gg L/\gamma$ (the rod subtends small angle) or when $v\ll c$ (retardation effects are generally negligible).

\par Equation (3.12) also indicates that apparent length depends on $d$. In fact if we look closely we will discover
that the smaller $d$ is, the more rapid contraction occurs near the observer. That is why the wheels of Gamow's
bicycle (see Fig. 2) contract non-uniformly and assume croissant-shape at one point . This observation also
suggests that a sphere under special conditions could appear convex as first pointed out by Scott and
Driel$^{[11]}$ and thoroughly discussed by Sufferin$^{[12]}$.

\section{Apparent superluminal velocities}
 \begin{figure*}[h]
 \centering
 \epsfig{file=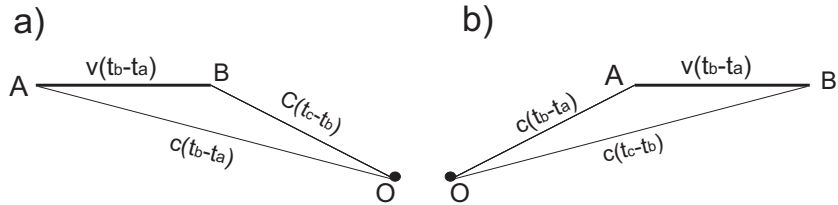, width=11cm}
 \caption{Geometrical explanation of superluminal velocities phenomenon.}
 \end{figure*}
As we previously noted the velocity of Gamow's bicycle in Fig. 2 exhibits fairly interesting behavior. The vehicle
clearly approaches the observer faster than recedes later, despite the fact that its factual velocity is constant.
The deceleration takes place in the vicinity of the closest approach to the observer. In order to understand this,
let us first consider a front light of an approaching bicycle in the stationary system $S$. The lamp emits a light
pulse at $t_{A}$ from the point $A$. The pulse reaches the observer in $t_{B}$ after travelling $c(t_{B}-t_{A})$.
At exactly the same time the front lamp emits another pulse from point $B$, after travelling the distance
$v(t_{B}-t_{A})$. The second pulse reaches the observer at $t_{C}$ after travelling the distance $c(t_{C}-t_{B})$.
According to the observer, the front light was in position $A$ at time $t_{B}$ and in $B$ at time $t_{C}$. The
observed velocity $u$ of the bike is, by definition, a ratio between the travelled distance to the time interval:
\begin{equation}
    u=\frac{v(t_{B}-t_{A})}{t_{C}-t_{B}}
\end{equation}
From Schwartz inequalities (see Fig. 5a) we obtain $c(t_{B}-t_{A})>c(t_{C}-t_{B})$, thus $u>v$. We can conduct
analogous considerations for the receding point (see Fig. 5b): $c(t_{C}-t_{B})>c(t_{B}-t_{A})$, thus $u<v$. This
is the reason why the apparent velocity of Gamow's bicycle is different from the real one and changes with time.
In fact we can calculate the velocity more precisely by simply differentiating (3.10), which is the equation of
apparent motion with respect to observer's time $t_{0}$:
\begin{equation}
    u=\frac{dx}{dt_{0}}=v\gamma^{2}(1-\frac{t_{0}v^{2} \gamma}{c\sqrt{d^{2}+(t_{0}v\gamma)^{2}}})
\end{equation}
A similar equation was first derived by Mathews and Lakshmann$^{[10]}$ who were also the first to discuss the
question of apparent velocities greater than the velocity of light in a physical journal. The expression for
apparent velocity (4.2) gives $u>v$ for approaching and $u<v$ for receding objects according to our expectations.
In order to closely analyze the dependance of $u$ on $d$ and $t_{0}$ we sketched the graph of $u(d,t)$ presented
in Fig. 6.
 \begin{figure*}[h]
 \centering
 \epsfig{file=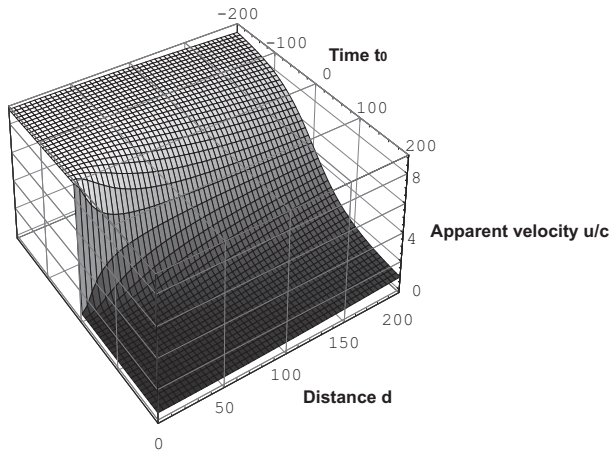, width=8cm}
 \caption{Apparent velocity as a function of observer's time and collision parameter.}
 \end{figure*}
Similarly to apparent length dynamics, the greater $v$ and smaller $d$ the more rapidly the apparent velocity
decreases, which resembles the acoustic Doppler effect. That is why different parts of Gamow's bicycle in Fig. 2
move relative to each other when approaching an observer. Eventually they assume the same velocity (when $t$ tends
to infinity). We can calculate limit velocities by putting $t_{0}\rightarrow \pm \infty$ in (4.2):
\begin{equation}
    u_{\pm\infty}=\frac{v}{1\pm\beta}.
\end{equation}
The plus sign in (4.3) refers to receding and minus to approaching bodies. The velocity $v>c/2$ ensures that the
apparent speed will exceed the speed of light! In turn, the maximal apparent velocity of a receding object is half
the speed of light.
\par We will now try to write down (4.2) in terms of apparent position of the body, which is far more useful in
practice. First let's define $\varphi$ as the angle between the velocity vector of the body and a line which links
the object and an observer. Let the distance to the body be $z$. In these terms we can write that:
$d=z\sin\varphi$ (collision parameter), $x=z\cos\varphi$ (apparent $x$ coordinate of the body) and
$vt_{0}=z(\cos\varphi+\beta)$ (calculated directly from (3.8)). If we substitute appropriate terms into (4.2) we
will obtain:
\begin{equation}
    u/c=\frac{\beta}{1-\beta\cos\varphi}.
\end{equation}
This expression could be derived directly. Let's consider a self-luminous object moving along a straight line with
velocity $v$. It emits light pulses at times $t_{1}$ and $t_{2}$. These impulses reach an observer at times
$t_{1}'$ and $t_{2}$': $t_{1}'=t_{1}+z/c$ and $t_{2}'=t_{2}+(z-v(t_{2}-t_{1}))\cos \varphi)/c$, where
$v(t_{2}-t_{1})$ is the distance travelled by the body and $v(t_{2}-t_{1})\cos \varphi$ tells us how much closer
the body is at the moment of the second emission (we assume both rays to be parallel since the interval $\Delta
t=(t_{2}-t_{1})$ is very small compared to $z/c$). Of course the apparent velocity is $u=v\Delta t/\Delta t'$. We
can easily calculate: $\Delta t'=\Delta t-\Delta t\beta\cos\varphi$ finally arriving with the expression (4.4).
Now we can see that when the body is seen at a right angle ($\varphi=\pi/2$ or $t_{0}=d/c$) the apparent velocity
equals the real one.
\par It is worthy to point out that apparent superluminal velocities have already been encountered in astrophysics.
First cases of so-called "superluminal extensions" were reported in early 70-s during observations of jets of two
quasars 3C273 and 3C279. There have been many reports in \textit{Science}$^{[25]}$ and \textit{Nature}$^{[26]}$
magazines since, concerning superluminal motions ranging from two up to thirty times the speed of light.
\section{Concluding Remarks}

It has been shown that Gamow's bicycle in Wonderland would in fact look differently than the description provided
by Gamow himself. It would appear grossly distorted and elongated (when approaching) or contracted (when
receding). Of course the picture seen by Mr. Thompkins would not fully correspond to Fig. 1 because of the
differences between human eyes and apparatus used to derive the mathematical description of the phenomenon. In
general Gamow's bicycle would be further distorted in the eyes of Mr. Thompkins and would be seen at a certain
angle. This would make the elongation (contraction) of the bicycle difficult to account for at greater times $\pm
t_{0}$. The velocity of the bicycle would also appear different from the factual one: it would approach Mr.
Thompkins with greater speed (under certain conditions even greater than $c$) and recede from him at smaller one
(with maximal being: $c/2$). However these are not the only differences between the reality and Gamow's
Wonderland. The spectrum and light intensity of both lamps (or in case of self-luminous bicycle - the entire
vehicle) will also be effected. When a self-luminous body moves with a certain velocity its spectrum changes due
to Doppler effect. The relation between the emitted $\lambda_{0}$ and observed $\lambda$ wavelengths is given by:
$\lambda=\lambda_{0}\gamma(1-\beta\cos\varphi)$. From section 4 we already know that
$\cos\varphi=x/\sqrt{x^{2}+y^{2}+d^{2}}$, so the Doppler shift of the emission point of the body depends on its
position. We can therefore expect that self-luminous bicycle would change its color when passing by the observer.
This aspect has already been thoroughly analyzed on the example of sphere$^{[18]}$ as well as the question of
light intensity change due to motion$^{[5][18][19][20]}$.
\par Knowing how a rapid motion affects the appearance and observable position of the body a question
arises: are we able to find out how the moving body really looks like, or is there any method to prove that
Lorentz contraction really occurs? The answer to both questions is yes. The simplest way to do that is to combine
measurement of the same moving body by two different researchers, observing the object at different angles. By
incorporating retardation effects into data analysis the contracted length and shape could be obtained. There is
also another way to experimentally determine the actual position and velocity of the body. It was developed by
Einstein and was given name of radar method. In brief, the method boils down to sending two electromagnetic
impulses towards the moving body and detecting their return after being reflected by the body. The desired
parameters could be obtained thanks to exact times of emission and observation of the reflected light.
Unfortunately all these methods are impractical when we deal with objects at astronomical distances. And it
appears this is the only place where retardation effects play a major role in observation process.
\section{Acknowledgments}
I would like to thank Dr Andrzej Dragan for having me interested in
the subject and for the invaluable discussions and support.



\begin{thebibliography}{}
 \bibitem[1] {} G. Gamow, "Mr. Thompkins in Wonderland" (Cambridge University Press, Cambridge 1958)
 \bibitem[2] {} A. Nowojewski, J.Kallas, A. Dragan, "On the Appearance of Moving Bodies", Am. Math. Month. \textbf{111} (11), 817 (2004)
 \bibitem[3] {} J. Terrell, "Invisibility of Lorentz Contraction", Phys. Rev. \textbf{116} (4), 1041 (1959)
 \bibitem[4] {} A. Lampa, "Wie erscheint nach der Relativitätstheorie ein bewegter Stab einem ruhenden Beobachter?" Z. Phys. \textbf{72}, 138(1924)(in German)
 \bibitem[5] {} V. F. Weisskopf, "The visual appearance of rapidly moving objects", Phys. Today \textbf{13} (9), 24 (1960)
 \bibitem[6] {} Sci. American \textbf{203} (1), 74 (1960)
 \bibitem[7] {} R. Penrose, "The apparent shape of relativistically moving sphere", Proc. Cambridge Phil. Soc. \textbf{55}, 137 (1959)
 \bibitem[8] {} M. Boas, "Apparent Shape of Large Objects at Relativistic Speeds", Am. J. Phys. \textbf{29} (5), 283 (1961)
 \bibitem[9] {} F.R. Hickey, "Two-dimensional appearance of a relativistic cube", Am. J.Phys. \textbf{47} (8), 711 (1979)
 \bibitem[10] {} P.M. Mathews, M. Lakshmann, "On the Apparent Visual Forms of Relativistically Moving Objects", Nuovo Cimento \textbf{12B} (1), 168 (1972)
 \bibitem[11] {} G. D. Scott, H. J. van Driel, "Geometrical Appearances at Relativistic Speeds", Am. J. Phys. \textbf{38}, 971 (1970)
 \bibitem[12] {} K. G. Suffern, "The apparent shape of a rapidly moving sphere", Am. J. Phys. \textbf{56} (8), 729 (1988)
 \bibitem[13] {} G. D. Scott, R. R. Viner, "The geomentrical appearence of large objects moving at relativistic speeds", Am. J.Phys. \textbf{33} (7), 534 (1965)
 \bibitem[14] {} A.P. French, "Special Relativity", Norton, New York 1968, p.149
 \bibitem[15] {} R. Resnick, "Introduction to Special Relativity", Wiley, New York 1968, p.77
 \bibitem[16] {} D. Hollenbach, "Appearance of a rapidly moving sphere: A problem for undergraduates", Am. J. Phys. \textbf{44} (1), 91 (1975)
 \bibitem[17] {} J. R. Burke, F. J. Strode, "Classroom exercises with the Terrell effect", Am. J. Phys. \textbf{59} (10), 912 (1991)
 \bibitem[18] {} U. Kraus, "Brightness and color of rapidly moving objects: The visual appearance of a large sphere revisited", Am. J. Phys. \textbf{68} (1), 56 (2000)
 \bibitem[19] {} J. M. McKinley, "Relativistic transformations of light power", Am. J. Phys. \textbf{47}, 602 (1979)
 \bibitem[20] {} J. M. McKinley, "Relativistic transformation of solid angle", Am. J. Phys. \textbf{48}, 612 (1980)
 \bibitem[21] {} R. Weinstein, "Observations of Length by a Single Observer", Am. J. Phys. \textbf{28} (7), 607 (1960)
 \bibitem[22] {} R. Bhandari, "Visual appearance of a Moving Vertical Line", Am. J.Phys. \textbf{38} (10), 1200 (1970)
 \bibitem[23] {} R. Bhandari, "Visual appearance of a moving vertical line revisited", Am. J.Phys. \textbf{46} (7), 760 (1978)
 \bibitem[24] {} D.W. Lang, "The Meter Stick in the Matchbox", Am. J.Phys. \textbf{38} (10), 1181 (1970)
 \bibitem[25] {} C.A. Knight \textit{et al}, "Quasars Revisited: Rapid Time Variations Observed Via Very-Long Baseline Interferometry", Science \textbf{173}, 225 (1971)
 \bibitem[26] {} R.T. Schilizzi, A.G. de Bruyn, "Large-scale radio structures of superluminal sources", Nature \textbf{303}, 26 (1983)
 \end{thebibliography}
\end{document}